\documentclass{ws-p9-75x6-50}

\begin{document}

\title{SEARCH FOR GLUONIC MESONS IN GLUON JETS}

\author{Peter Minkowski}

\address{Institute for Theoretical Physics,
           Univ. of Bern,  CH - 3012 Bern, Switzerland\\
E-mail: mink@itp.unibe.ch}

\author{\underline{Wolfgang Ochs}}

\address{Max Planck Institut f\"ur Physik, D - 80805 Munich, Germany\\
E-mail: wwo@mppmu.mpg.de}


\maketitle

\abstracts{
We present a short survey of the theoretical expectations on glueballs 
and hybrids as well as of the present phenomenological status.
The possibility to obtain new information on gluonic mesons 
from the study of gluon jets in comparison to quark jets is discussed.
}

\section{Introduction}\label{sec:intro}
The existence of glueballs, the bound states of two or more gluons, 
are an early expectation of QCD\cite{HFPM}. Calculations on a space time lattice
have  confirmed this expectation in the quantitative analysis. Their  
existence though is still in doubt despite continued efforts over the years.
The identification of the glueballs requires the analysis of production and
decay properties of the candidate mesons and has to be performed together
with the identification of the
lowest $q\overline q$ multiplets. Glueballs are searched for in 
particular in the ``gluon rich environment'' of central production 
in hadron hadron collisions (double Pomeron exchange), in radiative $J/\psi$
decays (hadrons from two intermediate gluons in the perturbative picture) 
or baryon antibaryon
annihilation into mesons near threshold. On the other hand, the production of
glueballs is expected to be suppressed in two photon processes.

An alternative possibility is the study of glueballs in gluon jets,
which -- although emphasized already  long ago\cite{pw} -- 
has not been systematically explored until now. Recent
discussions\cite{rs,mo1,sz} emphasize the interest in studying the
fragmentation region of the gluon jet. We will discuss here especially our
considerations\cite{mo1} how the detailed comparison of the fragmentation regions
of quark and gluon jets could provide new clues 
on the existence of gluonic mesons.

\section{Theoretical Expectations}\label{sec:theory}

There is general agreement in the theoretical analysis that the lightest
glueball should be the bound state of two valence gluons 
with antiparallel spins in a relative
S-wave with quantum numbers $J^{PC}=0^{++}$. The computation of its
mass in lattice QCD has been carried out with increasing
accuracy in recent years 
in the quenched approximation (without light sea quark-antiquark
pairs) and a recent summary\cite{cms} quotes the result
$m(0^{++})=1611$ MeV with a systematic error of about 10\%. First results
from unquenched calculations, i.e. including mixing effects with $q\overline q$
mesons, would shift the mass down 
by $\sim$20\% and so would lead to a mass of
$\sim$1300 MeV but these results may still be affected by large systematic
errors. 

For the hybrids the state of lowest mass is the ($q\overline qg$) 
bound state  with exotic quantum numbers  $J^{PC}=1^{-+}$ and the mass is
expected at $\sim$1900 MeV.

An alternative approach to hadron spectroscopy is provided by the 
QCD sum rules. Recent
calculations (see review\cite{narr}) for the $0^{++}$ glueball 
yield an estimated mass consistent with the quenched lattice result 
(with upper bound 2.16 $\pm$ 0.22 GeV), 
but in addition require a gluonic state near 1 GeV called $\sigma$
with strong mixing to  $q\overline q$ and a large width of 0.8 GeV. 

These results suggest that the lightest glueball  should be searched for
in the mass region 1.0 - 1.7 GeV, say. 

\section{Phenomenological Studies}

The lightest glueball should be found among the light scalar particles
or be mixed with them. The Particle Data Group (PDG\cite{PDG}) lists the
following $0^{++}$ states
\begin{equation}
f_0(400-1200)\ ({\rm or}\ \sigma) 
\quad f_0(980)\quad f_0(1370)\quad f_0(1500)\quad f_J(1710)\ (J=0) 
\label{scalars}
\end{equation}
Numerous suggestions on the interpretation of these states have been
proposed. We note two strategies\\ 
I) One starts from the quenched lattice result with a mass around 1600 MeV.\\
Then the $f_0(1500)$ is the natural candidate for the lightest glueball
or for the state with a strong admixture of gluonic component whereas the 
nearby states
$f_0(1370)$ and $ f_0(1710)$ should have a small gluonic admixture
as in an early proposal\cite{ac}.
However, rather different mixing schemes have been suggested 
depending in particular on the data included in the consideration (recent
review\cite{klempt}), for example, one with the largest gluon  component 
in  $ f_0(1710)$ and small one in $f_0(1500)$\cite{svw}. In these schemes the
$ f_0(980)$ and $ a_0(980)$ are often superfluous, they are taken as $K\overline K$
molecules and in this way removed from $q\overline q$ spectroscopy.\\
II) Alternatively, one can try to determine the scalar  $q\overline q$ 
nonet first 
considering the scalar particles below $\sim 1700$ MeV. Theoretical
considerations\cite{kmmp} as well as the phenomenological analysis\cite{mo} 
suggested
$f_0(1500)$ to be a member of the $q\overline q$ nonet with flavor mixing 
close to the octet. Especially the negative 
relative sign of the decay amplitudes into
$K\overline K$ and $\eta\eta$ determined in this analysis is not compatible
with a dominant glueball component of $f_0(1500)$.
The nonet can then be argued to be
\begin{equation}
q\overline q \ (0^{++}):\qquad
f_0(980) \quad a_0(980)\quad K^*(1430) \quad f_0(1500) 
\end{equation}
which is also in good agreement with the Gell Mann Okubo mass formula\cite{mo}.

The remaining light scalars in (\ref{scalars}), $f_0(400-1200)$ and $f_0(1370)$ 
are then interpreted as non-$q\overline q$ effects. One possibility
considered is their origin from $t$-channel Regge exchanges\cite{klempt};
in our view these two states are reflections of a single broad resonance, the
lightest binary glueball $gb(1000)$\cite{mo}. The arguments in favour of the
glueball hypothesis include: the strong central production in $pp$
collisions and the suppression in $\gamma\gamma$ collisions\cite{mo2};
against the glueball interpretation it has been put forward\cite{klempt} 
the lack of strong production in $J/\psi$ decay and 
certain inconsistencies in the decay ratios of  $f_0(1370)$
into $\sigma\sigma$ and $\rho\rho$ in different processes.

In the sector of hybrids there is now good evidence for states with the spin
exotic quantum numbers $J^{PC}=1^{-+}$ and masses of 1.4 and 1.6
GeV\cite{hybrids}. These masses are again lower than expected from latttice
calculations and one might discuss alternative interpretations of these
states, for example, four quark states.

To summarize the phenomenology, there are interesting candidates for gluonic
mesons but there is not yet a general consensus. It seems therefore
important to find further ways to identify the gluonic components 
of scalar mesons and in the remaining part of this outline we will 
discuss such a proposal.
  
\section{New Search Program for Gluonic Mesons}
Let us compare the production of hadrons in quark and gluon jets.
A fast hadron in a quark jet will likely carry the primary quark which 
initialized the jet
as a valence quark and this likelyhood will increase with increasing momentum
fraction $x$ of the hadron in the jet. This leads to the well established
phenomenology of quark fragmentation functions, for example, the $u$-quark
will produce more $\pi^+$ than $\pi^-$ of large momentum fraction $x$.
A natural extension of this phenomenology consists in supposing a similar
process for a gluon jet: particles with large momentum fraction $x$
should carry the initial gluon as valence gluon so the leading particles
should be glueballs or hybrids (see Fig. 1). 
So far, there is not yet a systematic
investigation of the gluon fragmentation beyond the study of single
inclusive spectra.\footnote{A negative result from a search for heavy
glueballs has been reported from a high statistics
analysis of the $K^0_s K^0_s$ mass spectra but no restriction
to the fragmentation region was applied.\cite{gluglue}}
We therefore propose investigating the gluon fragmentation in greater
detail and we discuss first the possible production mechanisms. 
\begin{figure}[ht]
\begin{center}
\mbox{\epsfig{file=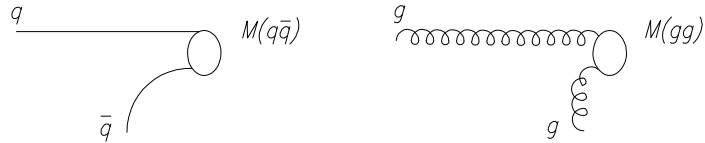,width=10cm,bbllx=1.0cm,bblly=17.0cm,bburx=18.cm,bbury=21.5cm}}
\end{center}
\caption{Fragmentation of a quark into a $q\overline q$ meson and of a gluon
into a glueball in the fragmentation region.}
  \label{fig:frag}
\end{figure}

\subsection{Hadron Production and Colour Neutralization}

An energetic  quark or gluon emerging from  a hard
collision process will generate a parton cascade
by subsequent gluon bremsstrahlung and quark pair
production.
The extension of such a cascade for a 100 GeV jet in space can be rather
large and exceed 100 f (see, for example \cite{dkmt,os}). The formation of
colour singlet systems should proceed during the evolution whenever
the separation of colour charges exceeds the confinement
length $R_c\sim 1$f. Two types of neutralization processes are possible
(see Fig. 2).
\begin{figure*}[t]
\begin{center}
\mbox{\epsfig{file=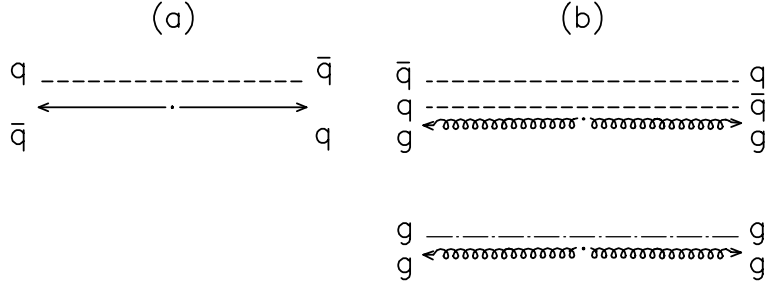,width=10cm}}
\end{center}
\caption[]{ 
The colour neutralization (a) of an initial $q\overline q$ pair
by $\overline q q$ and (b) of an initial $gg$ pair by either  double
colour triplet
$q\overline q$ or by colour octet $gg$}
\label{fig2}
\end{figure*}

a) Colour triplet neutralization.\\
Consider first 
the production of a 
$ q\overline q$ pair in a colour singlet state
corresponding to
separating colour triplet
charges  (possibly accompanied by secondary
 partons from bremsstrahlung).
The colour field between the primary quarks can be neutralized eventually by
the (nonperturbative)
 production of a soft quark antiquark pair.
 This process of triplet neutralization may repeat itself for the
various branchings inside the quark gluon cascade  until
the total energy is carried by
$q\overline q$ hadrons.
Also possible is the formation of  hybrid mesons from the gluons at the end
of the perturbative cascade and the soft  $q\overline q$.

b) Colour octet neutralization.\\
In case of a primary colour singlet $gg$ pair the field between the
separating colour octet charges can be neutralized at the confinement
distance  either by the production of a gluon pair or by the sequential
production of two quark pairs.
Only the mechanism with  gluon pair production could yield
pure gluonic bound states at the end.

It is not obvious to what extent the two types of neutralization mechanisms
are realized in a given
process. The octet neutralization could be an overall rare process
enhanced for particular kinematic configurations.
An experimental test has been proposed\cite{mo1} 
which does not rely on the existence of glueballs.

Consider the production of a hard gluon
which travels without gluon radiation for a while forming a jet with a
large rapidity gap empty of hadrons.
In this case the hard isolated gluon builds up an
octet field to the remaining partons, then the
octet neutralization mechanism
will become clearly visible if it exists. 
Namely, the total charge of
the leading particles beyond the gap should have total charge  $Q=0$;
on the other hand, the charge
distribution in case of double triplet neutralization should have 
in addition a component with charges $Q=\pm 1$. With increasing rapidity gap
the charge distribution should approach a limiting behaviour. 
An illustrative example of
the limiting charge distribution of the leading cluster in a jet is discussed 
elsewhere.\cite{mo1}

If the enhanced neutral component from octet neutralization
is not observed we would not expect a preferred source of glueballs
in gluon jets either. In this case glueballs would be produced through
their mixing with quarkonium states in any collision process
and not preferentially through their valence glue component.

\subsection{Search for Glueballs and Hybrids: Comparison of Quark and Gluon
Jets}

The following further study to identify the gluonic
component of hadrons is suggested.
One measures the mass distribution of the leading clusters with zero charge
($Q=0$), either clusters 
beyond a rapidity gap or clusters with large momentum fraction $x$ in the
jet.
We would expect a smaller background under a resonance
in the first case because of small
contributions from soft particles in the jet.
In any case the direct comparison of these mass spectra in quark and gluon
jets should reveal the possibly dominant component -- either gluonic or
quarkonic. The ratio of these spectra should reflect the global 
differences of both types of jets (different inclusive spectra,
phase space $\ldots$) and the local differences due to resonances of 
different intrinsic composition. 
Alternatively, one may also first normalize the cross
section of a state to a well known $q\overline q$ resonance nearby in mass  
and compare these ratios in quark and gluon jets.

\section{Summary}
We propose studying the fragmentation region of
quark and gluon jets with a selection of a rapidity gap or with
large momentum fraction $x$
of a particle cluster. The charge distribution of the leading cluster
should reveal the relevance of the color octet neutralization mechanism.
The relative production rates of the candidate gluonic
mesons,
in particular of the scalar mesons in (1), in the fragmentation region of 
quark and gluon jets should reveal their gluonic component.

An advantage of this method we consider the well established gluonic
nature of jets in certain configurations emerging from high energy 
collisions;
the gluonic nature of alternative processes like radiative decays
of Quarkonium or double Pomeron production is not yet established at the same
level of confidence.

\end{document}